\RequirePackage{fix-cm}
\documentclass[smallextended]{svjour3}       % onecolumn (second format)
\smartqed  % flush right qed marks, e.g. at end of proof
\usepackage{graphicx}
\begin{document}

\title{Local, Non-Geodesic, Timelike Currents in the Force-Free Magnetosphere of a Kerr Black Hole %\thanks{Grants or other notes
}

\titlerunning{Timelike Currents in the
Magnetosphere of a Kerr Black Hole
}        % if too long for running head

\author{Govind Menon        \and
        Charles Dermer %etc.
}


\institute{G. Menon \at
              Department of Physics and Chemistry, Troy University, Troy, Al, 36082, USA\\
              Tel.: +334-670-3924\\
              \email{gmenon@troy.edu}           %  \\
           \and
           C. Dermer \at
             Code 7653, Naval Research Laboratory, 4555 Overlook Ave., SW,  Washington DC 20375, USA
}

\date{Received: date / Accepted: date}

\maketitle

\begin{abstract}
In this paper, we use previously developed exact solutions to present some of the curious features of a force-free magnetosphere in a Kerr background. More precisely, we obtain a hitherto unseen timelike current in the force-free magnetosphere that does not flow along a geodesic. The electromagnetic field in this case happens to be magnetically dominated. Changing the sign of a single parameter in our solutions generates a spacelike current that creates an electromagnetic field that is electrically dominated.
\keywords{Black Hole Electrodynamics\and Force-free Magnetosphere}
 \PACS{PACS 04.20.Jb \and PACS 04.70.-s}
\end{abstract}

\section{Introduction}
\label{intro}
Electromagnetic fields evolving in a curved spacetime background have received much attention from physicists/astrophysicists alike for atleast four decades now.
In particular,  as a possible primary mechanism for the explanation of jets emanating from a rotating black hole, in  their classic paper by Blandford and Znajek \cite{BZ77}, they introduced the force-free, stationary, axisymmetric magnetosphere of a Kerr black hole. It was not  too long before Bekenstein and Oron \cite{BO77} generalized the setting  and introduced various conservation laws to enable calculations in general-relativistic  magnetohydrodynamics (GRMHD). Additionally, Thorne and MacDonald \cite{TM82} rewrote the covariant equations of electrodynamics in terms of  ``frame-dependent" electric and magnetic vector fields in order to bring in familiarity and intuition from classical electrodynamics.

Motivated by theoretical concerns, recent efforts in GRMHD include the work of Eric Gourgoulhon et. al. \cite{GMUE11}, in which they rewrite astrophysically relevant quantities in a coordinate independent formalism. In \cite{GJ14},  Gralla and Jacobson, present the current theoretical status of force-free electrodynamics which includes a concise discussion of all the known exact solutions. Astrophysical relevance of force-free electrodynamics has also been recently discussed by Okamoto \cite{IO06}.

 A class of exact solutions to the Blandford and Znajek set of equations describing the force-free magnetosphere of the Kerr spacetime were derived by Menon and Dermer in \cite{MD07}. This solution satisfied the  Znajek regularity condition at the event horizon (\cite{ZNA77}). This in turn made the solution well defined at the event horizon of the Kerr solution. While mathematically correct, the physical relevancy of the solution remaines elusive. Here, the current vector was proportional to the infalling principal null geodesic of the Kerr geometry. Naturally, the physical currents required a timelike decomposition of the net current vector. We presented one such possible decomposition in a subsequent paper (\cite{MD11}). While constructing this decomposition, we had a Blandford and Znajek energy extraction process in mind that was ultimately facilitated by a large scale Penrose type process. Here we were looking at extraction of energy using matter currents in addition to the electromagnetic Poynting flux. This required a net infalling current in the interior geometry and an outgoing current in the exterior geometry. To facilitate this,
we showed that for every infalling solution, there exists an outgoing solution. Both these solutions had the following properties: the net current was along a null geodesic vector, and the electromagnetic field was itself null in the sense that $\vec{B}^2-\vec{E}^2=0$.

There is a history of null current force-free solutions in various spacetime backgrounds. The flatspace monopole solution of \cite{Mic73}, and its generalization in a Schwarzschild background are all generated by a null current vector (\cite{GJ14}).  Motivated by our exact class of solutions and the solutions listed above, Brennan et. al. (\cite{BGJ13}) recently considered the force-free magnetosphere in a Kerr background when the current vector was proportional to the principal null geodesics of the Kerr geometry. What resulted was a sweeping generalization of all the solutions mentioned above. Their solution extended to the time dependent, non-axisymmetric case as well. Here too the current vector (by design) and the electromagnetic fields were  null  except for when in a Schwarzschild background a magnetic monopole was present; in which case the electromagnetic field was magnetically dominated in the sense that $\vec{B}^2-\vec{E}^2>0$.

In this paper, using a simple generalization of the previously obtained results, we will point out some of the curiosities of the force-free magnetosphere. We will continue to work in a Kerr background. It is not clear how  one can (uniquely) decompose the previously mentioned null current solutions into actual worldlines of charged particles. With this is mind, we seek to construct a current vector that is timelike in character. Here, we were only partially successful in that the class of solution are not valid everywhere even in the exterior geometry of the Kerr Black hole. As we shall show, this new solution is strictly not valid on the symmetry axis of the Black hole. However, this timelike force-free current has an unexpected feature in that it does not flow along a geodesic. In fact, the current vector does not even trace out a pre-geodesic.

 Our new solution is nothing more than a linear combination of the previously obtained infalling and out-going solutions. We begin our analysis by describing these solutions.

 %%%%%%%%%%%%%%%%%%%%%%%%%%%%%%%%%%%%%%%%%%%%
\section{The In-falling Solution }

%%%%%%%%%%%%%%%%%%%%%%%%%%%%%%%%%%%%%%%%%%%%%
The calculational details of this infalling solution is given in \cite{MD07}. Here, we simply state the results. Please note that the function $\Lambda$ that is used here is scaled differently from the original paper. They are related by the transformation:
$$\Lambda \frac{\cos \theta}{\sin^4 \theta}\rightarrow \Lambda\;.$$
The components of the electromagnetic fields in the Boyer-Lindquist coordinate system of the Kerr geometry are given by

\begin{equation}
E^{in} _\varphi=0=E^{in} _r \;,
\end{equation}
\begin{equation}
E^{in} _\theta =-\frac{2}{a^2}\;  \;\frac{\Lambda}{\sin\theta}\;,
\end{equation}
and
\begin{equation}
(B^{in})^\theta=0\;,
\end{equation}
\begin{equation}
(B^{in})^r=\alpha\;(H^{in})^r = \frac{2}{a} \; \;\frac{\Lambda\;\sin\theta}{\sqrt{\gamma}}\;,
\label{radb}
\end{equation}
and
\begin{equation}
\alpha\;B^{in} _\varphi = H^{in} _\varphi=\frac{2}{a^2}\; \Lambda\;.
\label{finalhphi}
\end{equation}
The definitions of the quantities listed above are explained in \cite{MD07} and \cite{MD05}.
The general subscripts $1,2,3$ in the Maxwell tensor corresponds to $r, \theta, \varphi$ respectively in our case. Here $\Lambda$ is an arbitrary function of $\theta$.
The current vector here is given by
\begin{equation}
I^{in} =-\frac{2}{a^2\; \alpha\; \sqrt\gamma} \;\;\frac{d\Lambda }{d \theta}\;n\;,
\end{equation}
where $n$ is the infalling principle null geodesic of the Kerr geometry:
\begin{equation}
n= \frac{r^2+a^2}{\Delta}\;\partial_t\; - \;\partial_r \;+\;\frac{a}{\Delta}\;\partial_\varphi.
\end{equation}
The superscript ``{\it in}" refers to the fact the these field quantities describe an infalling current. Transforming the Maxwell tensor $F^{in}_{\mu \nu}$ into the Kerr-Schild coordinate system, we see that
\begin{equation}
F^{in}_{\bar t \bar r} = F^{in}_{\bar t \bar \varphi}=  F^{in}_{\bar r \bar \varphi}=F^{in}_{\bar r \bar \theta}=0 \;,
\end{equation}
\begin{equation}
F^{in}_{\bar t \bar \theta} = - E^{in}_\theta\;,
\end{equation}
\begin{equation}
F^{in}_{\bar \theta \bar \varphi} = \sqrt{\gamma}\;(B^{in})^r\;,
\end{equation}
and
$$I^{in} =\frac{2}{a^2\; \alpha \;\sqrt\gamma} \;\;\frac{d\Lambda }{d \theta}\;\partial_{\bar r}\;.$$
Thus we see that the fields and currents are well defined on the event horizon $r = r_+$ as well. This is necessarily so since we had insisted on the Znajek regularity condition given by eq.(\ref{znaregcond}) in the derivation of our solution (\cite{MD07}).

%%%%%%%%%%%%%%%%%%%%%%%%%%%%%%%%%%%%%%%%%%%%
\section{The Out-going  Solution}
%%%%%%%%%%%%%%%%%%%%%%%%%%%%%%%%%%%%%%
In this case, the electromagnetic field quantities are given by
\begin{equation}
E^{out} _\varphi = 0 =  E^{out} _r \;,
\end{equation}
\begin{equation}
 E_\theta ^{out} = - \frac{2}{a^2}\;  \;\frac{\tilde\Lambda}{\sin\theta}\;,
\end{equation}

\begin{equation}
 (B^{out})^\theta =0\;.
\end{equation}
\begin{equation}
 (B^{out})^r=\alpha\;( H^{out})^r = \frac{2}{a} \; \;\frac{\tilde \Lambda \sin\theta}{\sqrt{\gamma}}\;,
\end{equation}
and
\begin{equation}
\alpha\; B_\varphi ^{out}=  H_\varphi ^{out}= - \frac{2}{a^2}\; \tilde\Lambda\;.
\label{finalhhphi}
\end{equation}
Notice the extra minus sign in the expression for $B_\varphi ^{out}$. This small change is sufficient to modify the current vector to
\begin{equation}
I^{out} =- \frac{2}{a^2 \alpha \sqrt\gamma}\; \frac{d\tilde \Lambda }{d \theta}\;l\;.
\end{equation}
Here $l$ is the principle outgoing null geodesic of the Kerr geometry given by
\begin{equation}
l = \frac{r^2+a^2}{\Delta}\;\partial_t +  \partial_r + \frac{a}{\Delta}\partial_\varphi\;.
\end{equation}
The out-going solution allows for the extraction of energy  via the electromagnetic Poynting flux from the black hole.
Here the rate of energy extraction is given by

%$$\frac{d \tilde {\cal E}_{EM}}{dt}=-\int \tilde H_\varphi \tilde \Omega \tilde B^r \sqrt {\gamma_{rr}} dA$$
\begin{equation}
  \dot {\cal E}_{EM}^{out} = \frac{8\pi}{a^4} \int_{0}^{\pi} \frac{\tilde \Lambda^2\; }{\sin\theta} \;d\theta \geq\;0\;.
\label{engext}
\end{equation}
Transforming the Maxwell tensor $F^{out}_{\mu \nu}$ into the Kerr-Schild coordinate system, we see that
\begin{equation}
F^{out}_{\bar t \bar r} = F^{out}_{\bar t \bar \varphi}=  F^{out}_{\bar r \bar \varphi}=0 \;,
\end{equation}
\begin{equation}
F^{out}_{\bar t \bar \theta} = - E^{out}_\theta\;,
\end{equation}
\begin{equation}
F^{out}_{\bar \theta \bar \varphi} = \sqrt{\gamma}\;(B^{out})^r\;.
\end{equation}
Unlike in the infalling case, however, since these solution do not satisfy the Znajek regularity condition, here $F^{out}_{\bar r \bar \theta} \neq 0$, instead
$$F^{out}_{\bar r \bar \theta} = \frac{-4 \rho^2 \tilde \Lambda}{a^2 \sin \theta \Delta}\;.$$
Clearly $F^{out}_{\bar r \bar \theta}$ is not valid at the horizon.

%%%%%%%%%%%%%%%%%%%%%%%%%%%%%%%%%%%%%%%%%%%%%%%%%%
\section{Timelike Currents In The Force-Free Magnetosphere}
%%%%%%%%%%%%%%%%%%%%%%%%%%%%%%%%%%%%%%%%%%%%%%%%%%%%
Since Maxwell's equations are linear in a fixed curved background, we create a new solution by forming a linear combination of the two solutions given above in hopes of creating a timelike current in the black hole magnetosphere.
Accordingly, we now set
\begin{equation}
 E_\varphi=0= E_r \;,
\end{equation}
\begin{equation}
E_\theta =-\frac{2}{a^2}\;  \;\frac{(\Lambda + \tilde \Lambda)}{\sin\theta}\;,
\end{equation}
and
\begin{equation}
 B^\theta=0\;,
\end{equation}
\begin{equation}
B^r=\alpha\;H^r = \frac{2}{a} \; \;\frac{(\Lambda + \tilde \Lambda)\sin\theta}{\sqrt{\gamma}\;}\;,
\end{equation}
and
\begin{equation}
\alpha\;B_\varphi = H_\varphi=\frac{2}{a^2}\; (\Lambda - \tilde \Lambda)\;.
\end{equation}
This is consistent with the current 4 vector
\begin{equation}
I = -\frac{2}{a^2\; \alpha\; \sqrt\gamma}\; \left [ \frac{d\tilde \Lambda}{d \theta}\; l +  \frac{d\Lambda}{d \theta}\; n \right].
\label{lincombcurrent}
\end{equation}
Since the only non-vanishing component of the electric field is $E_\theta$, once again from the above expression for the current density vector, we find that
$$E \cdot J =0\;.$$
The other requirement for a force-free magnetosphere is given by
$$\rho_c E + J \times B = 0\;.$$
In our case we have that
$$\rho_c E + J \times B $$
$$= (\rho_c ^{in} +  \rho_c ^{out} ) ( E^{in} +  E^{out}) + (J^{in} + J^{out}) \times ( B^{in} +  B^{out})$$
$$= \rho_c^{in}  E^{out} +  \rho_c^{out} E^{in}   + J^{in}  \times  B^{out} +  J^{out} \times  B^{in}=0\;. $$
The above expression for the force-free condition is only non-trivial along the $\theta$ component. Therefore, we now require that
$$\rho_c^{in}  E^{out} _\theta    -\sqrt{\gamma} [ (J^{in})^r  \times  (B^{out})^\varphi -  (J^{in})^\varphi \times  (B^{out})^r] $$
$$+ \;\rho_c^{out} E^{in} _\theta   -\sqrt{\gamma} [ (J^{out})^r  \times  (B^{in})^\varphi -  (J^{out})^\varphi \times  (B^{in})^r]=0\;. $$
Substituting in expressions for the fields and currents, we get
\begin{equation}\frac{8 \rho^2}{a^4  \Delta \sin\theta \sqrt{\gamma}} \left(\frac{d\Lambda}{d \theta}\; \tilde \Lambda + \frac{d\tilde \Lambda}{d \theta} \; \Lambda\right) =0\;.
\label{cirscrosseq}
\end{equation}
This reduces to
$$ \frac{d }{d \theta}\Big (\Lambda\;\tilde\Lambda\Big ) = 0$$
which is easily solved to give
\begin{equation}
\Lambda \tilde \Lambda =k \;.
\label{tildelamsolved}
\end{equation}
Here $k$ is an integration constant. Obviously, when $k=0$ we revert to one of the previous cases.
From eq.(\ref{lincombcurrent}) we get that
$$I^2 = \frac{-16}{a^4 \rho^2 \sin^2 \theta \Delta}\;\frac{d\tilde \Lambda}{d \theta}\;\frac{d \Lambda}{d \theta} \;.$$
The force-free condition given by eq. (\ref{cirscrosseq}) implies that
$$\frac{d\tilde \Lambda}{d \theta}\;\frac{d \Lambda}{d \theta} = \frac{-k}{\Lambda^2}\left(\frac{d \Lambda}{d \theta} \right)^2\;.$$
Therefore, we get that
$$I^2 = \frac{16 k}{a^4 \rho^2 \sin^2 \theta \Delta}\;\left(\frac{1}{\Lambda}\frac{d\tilde \Lambda}{d \theta}\right)^2\;.$$
$k >0$ makes the current density vector given by eq.(\ref{lincombcurrent}) spacelike, however, when $k <0$, we have a timelike current vector field, and it now becomes a candidate for actual charge carrying particles. It is important to note that eq.(\ref{lincombcurrent}) is regular at the event horizon if and only if $\tilde \Lambda = 0$. I.e., the generalizations we are now constructing is valid only in the exterior geometry. Eq.(\ref{lincombcurrent}) is conveniently rewritten as
\begin{equation}
I = -\frac{2}{a^2\; \alpha\; \sqrt\gamma}\;  \frac{d \Lambda}{d \theta}\; \sqrt{\frac{-4\; k \;\rho^2 }{\Lambda^2 \;\Delta}}\; u \;,
\end{equation}
where
\begin{equation} u= \sqrt{\frac{\Lambda^2 \Delta}{-4 \;k \rho^2}}\;\left(n -\frac{ k}{ \Lambda^2}\;l \right)
\label{timecur}
\end{equation}
is the normalized timelike vector field when $k<0$.
In this case, $u$ is also future pointing.
$u$ outgoing when
$$\Lambda^2 < -k \;.$$
Despite being force-free, it is important to note that these currents are not flowing through geodesic curves (or even pre-geodesic curves):
$$\nabla_u u =-u \left[\ln \sqrt{\frac{\Lambda^2 \Delta}{-4 \;k \rho^2}} \;\right] \;u + \frac{\Delta}{4 \rho^2}\;\;\Big[\nabla_n l+\nabla_n l\Big] $$

$$ =-\frac{1}{2}u\left[\ln \frac{ \Delta}{ \rho^2} \;\right] \;u\;
-\frac{1}{ \rho^4}\Big[\left(M a^2  \cos^2 \theta+r a^2\sin^2\theta - M r^2\right)\partial_r $$$$+ a^2 \cos\theta \sin\theta \;\partial_\theta\Big]
\;. $$
There is no mathematical inconsistency here since force-free simply means that $F_{\mu\nu}\; I^\nu = 0$. Nevertheless, the existence such future pointing, albeit local (as we shall see) timelike force-free currents that do not flow along geodesics is at the very least curious. Perhaps the true physical content of the force-free condition for a fluid is still unclear within the context of a curved spacetime.

The magnetosphere here is indeed magnetically dominated. A quick calculation reveals that
\begin{equation}B^2 - E^2 = -8 k \;\left[\frac{\rho^2 (\Sigma^2- \Delta a^2 \sin^2\theta )+\Delta \Sigma^2}{a^4\sin^2\theta\; \Sigma^2 \;\Delta\; \rho^2}\right]\;.
\label{emstrength}
\end{equation}
Note that $-k > 0$, and for $r > r_+ > a$ we have that $\Sigma^2-\Delta a^2 \sin^2\theta > 0$.
The radial component of the magnetic field, at large distances does indeed fall off like a monopole term:
$$
B^r \approx \frac{2}{a} \; \;\frac{(\Lambda + \tilde \Lambda)}{r^2}\;.
$$
But this we have come to expect in spacetime with singularities.
%%%%%%%%%%%%%%%%%%%%%%%%%%%%%%%%%%%%%%%%%%%
\section{Calculation At The Poles}
%%%%%%%%%%%%%%%%%%%%%%%%%%%%%%%%%%%%%%%%%%%%%%%
We have hinted that our solution might have difficulties along the symmetry axis of the black hole. We shall clarify this point in this section. The transformation of the Maxwell tensor from the Kerr-Schild coordinates to a coordinate system that is valid along the poles can be accomplished using the results from section \ref{KSA}. Upon completing the necessary transformations, one finds that barring accidental cancelations in specific solutions, if the Maxwell tensor has to be well defined along the symmetry axis of the external Kerr geometry, then
$F_{\bar t \bar r},F_{\bar t \bar \theta}, F_{\bar r \bar \theta}$ and
$F_{\bar t \bar \varphi}/\sin \theta,F_{\bar r \bar \varphi}/\sin \theta, F_{\bar \theta \bar \varphi}/\sin \theta$
have to be well defined along the poles ($\bar \theta = 0, \pi$).
In our case, this becomes problematic since the above requirements implies the finiteness of
$\Lambda+ \tilde \Lambda, (\Lambda+ \tilde \Lambda)/\sin\theta, \tilde \Lambda/\sin\theta$ at the poles.
In particular, we must have that $$\lim_{\theta \rightarrow 0, \pi} \tilde \Lambda =0\;.$$
This is certainly impossible considering eq. (\ref{tildelamsolved}) and the timelike vector field in eq. (\ref{timecur}). While there is a formal divergence in the components of the Maxwell tensor along the poles, the observable quantities like the rate of energy and angular momentum extraction can remain finite for suitable choices of $\Lambda$.
In particular setting $k= 1$ and $\Lambda^2 = \cos \theta/2$ when $0 \leq \theta \leq \pi/2$ and  $\Lambda^2 = \cos (\pi/2 -\theta/2)$ when $\pi/2\leq \theta \leq \pi$ gives the rate of energy extraction via the electromagnetic Poynting flux by
$$\dot {\cal E} = \frac{16 \pi (\sqrt{2}-1)}{a^4}\;,$$
and the rate of angular momentum extraction is given by
$$\dot {\cal L} = \frac{16 \pi }{3 a^3}(5 \sqrt{2}-8)\;.$$

%%%%%%%%%%%%%%%%%%%%%%%%%%%%%%%%%%%%%%%%%%%
\section{Conclusion}
%%%%%%%%%%%%%%%%%%%%%%%%%%%%%%%%%%%%%%%%%%%%%%%

The search for exact solutions to the equations of force-free electrodynamics in the magnetosphere of a rotating black hole is important because the existence of the radio-loud class of active galactic nuclei (AGN) with jets could be explained if they are powered by the energy from supermassive black-hole rotation in addition to accretion (\cite{BL90}). 
Analytic progress has been slow following the original Blandford-Znajek (\cite{BZ77}) solution that generalized the flat-spacetime split monopole solution of Michel (1973) to a Kerr spacetime.
We have focused on time-independent and axisymmetric solutions to the constraint equation of force-free electrodynamics, and obtained a pair of approximate ($O(1/r^2)$) solutions that satisfies the Znajek regularity condition at the event horizon (Menon \& Dermer \cite{MD05}). One of them generalizes the split-monopole solution to all values of $a < 1$, though with most energy dissipated along the equatorial plane. The other solution feeds energy into the system.
In later work (Menon  \& Dermer \cite{MD07} , \cite{MD11}, \cite{DMbook}), we constructed a class of exact solutions for this force-free system in a Kerr background wherein the net current vector flowed through the principal null geodesics of the Kerr spacetime. Using this null geodesic as an ansatz, these solutions were subsequently generalized by Brennan et al. (\cite{BGJ13}), who obtained a large class of exact, time-dependent, non-axisymmetric solutions. Additional attempts to solve the equations include those of Beskin et al. (\cite{BIP92}), who considers an accretion disk along the equator, and Contopoulos et al. (\cite{CKP13}), who treat a pulsar-like solution to the Grad-Shafranov equation with current flowing through the light cylinder based. Lyutikov (\cite{LY11}) also uses the Grad-Sharfranov equation to solve for the time evolution of a pulsar magnetosphere to the monopole solution as the pulsar collapses to a black hole.

The (modest) catalog of previously existing analytical solutions to the force-free magnetosphere in a Kerr background have featured two consistent properties: the net current vector is null; and the electromagnetic field is null as well, by which we mean that $B^2-E^2 = 0$.  Here we focused on relaxing exactly these two above mentioned, albeit un-imposed restrictions. We have been successful in creating a net timelike current that does not flow along a geodesic even in a force-free magnetosphere. As expected, the electromagnetic field here is magnetically dominated, and the magnetic field plays a major role in the particle motions, possibly accounting for the existence of timelike currents that do not follow geodesics. Here, the current vector can be identified with the trajectories of charged particles in the magnetosphere.
A spacelike current is equally easy to obtain by setting $k > 0$. From eq. (32), we see that the electromagnetic field becomes electrically dominated in the presence of a spacelike net current vector.

The physical utility of these new solutions is unclear because these solutions do not satisfy the Znajek regularity condition (\cite{ZNA77}) at the event horizon.  Furthermore, the solution diverges along the rotation axis of the black hole. An astrophysically relevant solution may be obtained by restricting the solutions to the regions where it is not unphysical, or by stipulating the existence of a current sheet at the event horizon that can be matched to the exterior solution.  The role of the accretion-disk surrounding the black hole can further support currents that could participate in the global current system and energy release.  Nevertheless, because $u$ is timelike and outward pointing, and the current given by eq. (30) allows energy extraction from the black-hole via the Poynting flux, these results may provide a basis for constructing a physically valid jetted solution that is regular at the event horizon, has current flows that could be dictated by accretion-disk conditions, and produces a collimated  plasma jet. This is the subject of current research by the authors.

\section*{Acknowledgments}
The first author would like to thank Troy University for their continued support of our research in black hole astrophysics. The Chief of Naval research funds and supports the second author.

\appendix
\section{Kerr Geometry Essentials}
\label{KGE}
For completeness, we define the various Kerr coordinates used. For asymptotic analysis, the Boyer-Lindquist coordinates are preferred, while the horizon and the interior region ($r \leq r_+$) are analyzed using the usual Kerr-Schild coordinate system. The symmetry axis is analyzed using a Cartesian type coordinate system.
\subsection{ Boyer-Lindquist Coordinates}
In the Boyer-Lindquist
coordinates $\{t,r,\theta,\varphi\}$ of the Kerr geometry, the metric takes the form
$$
ds^2=( \beta^2 - \alpha^2 )\;dt^2 \;+  \;2 \;\beta_\varphi \;d\varphi
\;dt
+\gamma_{rr}\; dr^2 + \;\gamma_{\theta \theta}\; d\theta^2 +
\;\gamma_{\varphi\varphi}\;d\varphi^2 \;,$$
where the metric coefficients are given by
$$\beta^2-\alpha^2 \;= \;g_{tt} \;=\; -1 + \frac{2Mr}{\rho^2}\;,\;\;\beta_\varphi \; \equiv g_{t \varphi}\; = \;\frac{-2Mr a
\sin^2\theta}{\rho^2}\;,\;\;\;\gamma_{rr} =
\frac{\rho^2}{\Delta}\;,$$
$$
\gamma_{\theta \theta} = \rho^2, \;\; {\rm and}\;\; \gamma_{\varphi \varphi} = \frac{\Sigma^2 \sin^2\theta}{\rho^2}\;.
$$
Here,
$$\rho^2 = r^2 + a^2
\cos^2\theta\;,\;\;\;\Delta = r^2 -2 M r + a^2$$
and
$$
\Sigma^2 = (r^2 + a^2)^2 -\Delta \; a^2 \sin^2\theta\;.
$$
 Additionally
$$
\alpha^2 = \frac{\rho^2 \Delta}{\Sigma^2}, \;\;\; \beta^2 =
\frac{\beta_\varphi^2}{\gamma_{\varphi \varphi}}\;,\;\sqrt{\gamma} = \sqrt{\frac{\rho^2\;\Sigma^2}{\Delta}}\;\sin \theta\;,$$
and
$$\sqrt{-g}=\alpha\; \sqrt{\gamma} = \rho^2 \sin\theta\;.$$
The parameters $M$ and $a$ are the mass and angular momentum per unit
mass respectively of the Kerr black hole. The horizons $H_\pm$  are located at   $ r_\pm = M \pm \sqrt{M^2 - a^2} $.
\subsection{ Kerr-Schild Coordinates}
Kerr-Schild coordinates are given by the transformation

\begin{equation}
\left[\begin{array}{c}
d\bar t\\
d \bar r\\
d \bar \theta\\
d \bar \varphi\\
\end{array}\right]
=\left[\begin{array}{cccc}
1 & G& 0& 0\\
0& 1& 0& 0\\
0 & 0 & 1 & 0\\
0& H& 0& 1\\
\end{array}\right]
\left[\begin{array}{c}
d t\\
d r\\
d \theta\\
d \varphi\\
\end{array}\right],
\label{transformup}
\end{equation}
where
\begin{equation}
G = \frac{r^2+a^2}{\Delta}\;\;\;\;\; {\rm and}\;\;\;\;\; H = \frac{a}{\Delta}\;.
\end{equation}
In this frame, the metric becomes
\begin{equation}
 g_{ \mu  \nu} = \left[\begin{array}{cccc}
z-1& 1& 0& -za\sin^2\theta\\
1& 0& 0& -a\sin^2\theta\\
0 & 0& \rho^2 & 0\\
-za\sin^2\theta & -a\sin^2\theta& 0& \Sigma^2 \sin^2\theta/\rho^2\\
\end{array}\right] \;,
\label{kerrschildmetric}
\end{equation}
where $z = 2Mr/\rho^2$. We pick our time orientation for the Kerr geometry such that the null vector field $-\partial_{\bar r}$ is future pointing everywhere.

\subsection{The Kerr Symmetry-Axis}
\label{KSA}
For the case of Minkowski spacetime, the polar singularity of the
spherical coordinate system is removed by going to the cartesian
coordinate system (although usually, the cartesian coordinate system
is the natural starting point). The coordinate singularity along the Kerr poles can similarly be eliminated by a ``cartesian" like coordinate system as well. The new coordinates are labeled
$T, X, Y, Z$.
This is accomplished by
the following transformations:
$$X = (\bar r \cos \bar \varphi -  a \sin \bar \varphi)\; \sin \bar \theta\;,\;Y = (\bar r \sin \bar \varphi + a \cos \bar \varphi) \;\sin \bar \theta\;,\;Z = \bar r \;\cos \bar \theta$$
and $T = \bar t - \bar r\;.$
\section{ Equations Of Electrodynamics In Stationary Spacetimes}
\label{EESS}

We only state the relevant equations of electrodynamics of stationary spacetimes. For a detail development, see \cite{DMbook}. Maxwell's equations can be written as
\begin{equation}
\nabla_\beta \star \;F^{\alpha \beta} = 0 \;, \;{\rm and} \;
\nabla_\beta  F^{\alpha \beta} = I^\alpha\;.
\label{maxeq}
\end{equation}
Here $F^{\alpha \beta}$ is the Maxwell stress tensor,  $I^\alpha$ is the four vector of the electric current  and $\nabla$ is the covariant derivative of the geometry. $\star \;F$ is
the two form defined by
\begin{equation}
\star \;F^{\alpha \beta} \equiv \frac{1}{2}\epsilon^{\alpha \beta \mu \nu} F_{\mu \nu}\;.
\end{equation}
Here, $\epsilon_{\alpha \beta \mu \nu}$ is the completely antisymmetric Levi-Civita tensor density of spacetime such that $
\epsilon_{0123}= \sqrt{-g}=  \alpha \sqrt{\gamma}
$. In the 3+1 formalism, where $\partial_0$ is the asymptotically stationary timelike killing vector field, $ E$ and $ B$ are defined so that
\begin{equation}
 F_{ \mu  \nu} =
\left[\begin{array}{cccc}
0&  -E_1&  -E_2 & -E_3\\
  E_1 & 0 & \sqrt{\gamma}\; B^3 & - \sqrt{\gamma}\; B^2\\
E_2& - \sqrt{\gamma} \;B^3 & 0 & \sqrt{\gamma} \;B^1\\
E_3 & \sqrt{\gamma}\;B^2 & - \sqrt{\gamma} \;B^1 & 0\\
\end{array}\right] \;.
\label{fdown}
\end{equation}
We also define dual vectors $ D$ and $ H$ by
\begin{equation}
 * F_{ \mu  \nu} =
\left[\begin{array}{cccc}
0&  H_1&  H_2 & H_3\\
-  H_1 & 0 & \sqrt{\hat\gamma}\; D^3 & - \sqrt{\hat\gamma}\; D^2\\
- H_2& - \sqrt{\hat\gamma}\; D^3 & 0 & \sqrt{\hat\gamma} \;D^1\\
- H_3 & \sqrt{\hat\gamma} \; D^2 & - \sqrt{\hat\gamma} \; D^1 & 0\\
\end{array}\right] \;.
\label{starfdown}
\end{equation}
Naturally, $F$ and $\star \;F$ are not independent. They are related by
\begin{equation}
\alpha D = E- \beta \times B
\label{contitutive1}
\end{equation}
and
\begin{equation}
H = \alpha B - \beta \times D\;.
\label{contitutive2}
\end{equation}
Here,
\begin{equation}
(A \times B)^i \equiv \; \epsilon^{ijk}  \; A_j\; B_k\;,
\end{equation}
where $\epsilon^{ijk}$ is the Levi-civita tensor of our absolute space defined $x^0 = {\rm constant}$. Also, $\beta$ is the shift dual vector given by $\beta = \beta_\varphi \;d\varphi$. Naturally, the spatial coordiantes are given by $(x^1, x^2, x^3)$, and three vectors $ E, B, D, H$ live in this absolute space. Now, Maxwell's equations can be re-written as
\begin{equation}
\tilde \nabla \cdot B = 0\;,
\label{divb}
\end{equation}
\begin{equation}
\partial_t B + \tilde \nabla \times E = 0\;,
\label{faraday}
\end{equation}
\begin{equation}
\tilde \nabla \cdot D = \rho_c\;,
\label{maxcharge}
\end{equation}
and
\begin{equation}
-\partial_t D + \tilde \nabla \times H = J\;,
\label{maxcurrent}
\end{equation}
where $\rho_c = \alpha I^t$ and $J^k = \alpha I^k$. Here $\rho_c$ is the charge density and $J$ is the electric 3-current. $\tilde \nabla$ is the covariant of the 3 space with the induced metric.
The force-free condition that we will enforce is
\begin{equation}
F_{\nu \alpha}\; I^\alpha =0\; .
\label{divtemforce}
\end{equation}
This condition takes the form
\begin{equation}
E \cdot J = 0
\label{fofree1}
\end{equation}
and
\begin{equation}
\rho_c E + J \times B = 0.
\label{fofree2}
\end{equation}
For the case of a stationary,  axis-symmetric,  force-free magnetosphere, it is easy to show that there exists $\omega = \Omega\;\partial_{\varphi}$ such that
\begin{equation}E = - \omega \times B\;.\label{omdefeqn}\end{equation}
Additionally,  \cite{ZNA77} showed that
\begin{equation}
H_\varphi \left |_{r_+} = \frac{\sin^2\theta}{\alpha}\; B^r\; (2Mr\; \Omega -a) \right |_{r_+}
\label{znaregcond}
\end{equation}
is the required condition in the Boyer-Lindquist coordinates that the otherwise bounded fields must satisfy so that they continue to be well defined in the Kerr-Schild coordinates at the event horizon. Eq.(\ref{znaregcond}) is referred to as the Znajek regularity condition. The rate of electromagnetic extraction of energy from the force free magnetosphere is given by
\begin{equation}
\frac{d  {\cal E}_{EM}}{dt}=\int_{r_+}  S^r \sqrt {\gamma_{rr}} \;dA = \int_{r_+}   S^r \;\sqrt{\gamma}d\theta\;d\varphi\;,
\label{engextractformula}
\end{equation}
where $S^i = (-\alpha \; T^i~_t)$.
Using the Znajek regularity condition we get that
$$\frac{d  {\cal E}_{EM}}{dt}=-2\;\pi \int_{r_+}E_\theta ^2\;\frac{(2Mr\; \Omega -a)}{\rho^2\;\Omega}\;\sin \theta \; d\theta\;,$$
which is indeed positive when
\begin{equation}
0< \Omega < \Omega_H\;.
\label{omrange}
\end{equation}
%\section{Section title}
%\label{sec:1}
%\subsection{Subsection title}
%\label{sec:2}
%\paragraph{Paragraph headings} Use paragraph headings as needed.
%\begin{figure}
%  \includegraphics{example.eps}
%\end{figure}
%

%\begin{acknowledgements}
%\end{acknowledgements}



\end{document}